\documentstyle[12pt,psfig]{article}
\begin{document}
\begin{titlepage}
\flushright{DTP/96/23}
\vspace{1cm}
\center{\Large {\bf A Local Model of Explicit Wavefunction Collapse}}
\vspace{1cm}
\center{\large Chris Dove\footnote{E-mail:C.J.Dove@durham.ac.uk}} \\ and \\
{\large Euan J. Squires\footnote{E-mail:E.J.Squires@durham.ac.uk}}
\vspace{0.2cm}
{\it\center{Department of Mathematical Sciences \\ University of Durham
\\ Durham City, DH1 3LE, England.}}
\vspace{0.2cm}
\center{May 1996}
\vspace{0.5cm}
\begin{abstract}
A model of spontaneous wavefunction collapse, which is explicitly local and
Lorentz-invariant, is defined. Some of the predictions of the model for
specific experimental situations are derived. It is shown that, although
incompatible collapses, e.g. on opposite sides of an EPR-type of
experiment, can occur, they will not persist in time and that
eventually only compatible results will be obtained. The probabilities of
particular results, however, will in general not agree with the predictions
of quantum theory. We argue that it is unlikely that the deviations would
have been seen in any experiment yet performed.
\end{abstract}
\end{titlepage}
\section{Introduction}
In 1986 Ghirardi, Rimini and Weber \cite{GRW}
showed that it was possible to construct a realistic model 
describing explicit wavefunction collapse in such a way that, in many
situations, the correct
predictions of quantum theory were maintained but real experiments actually
had results. Their work has since been developed in a number of ways
\cite{GPR} and it 
is generally agreed that it provides a satisfactory resolution of the
measurement problem of quantum theory, at least in the non-relativistic
domain. As originally presented, however, the model was clearly non-local
and not Lorentz invariant. Recently, attempts have been made to develop
versions of the collapse models which, whilst retaining the non-locality,
are nevertheless Lorentz invariant\cite{PP1,PP2,GGP}.
Perhaps the best one can say of these 
models is that they are partially successful. They certainly raise several
interesting issues.

In this work we shall take a different approach and endeavour to construct
a {\it local}, and Lorentz invariant version of the collapse model.
We know of course that this cannot agree in all respects with the 
predictions of orthodox quantum theory, and it is one object of this work to 
see where the disagreement lies and whether it is detectable.
Note that even the original GRW model does not completely agree with 
quantum theory, and this requires severe constraints to be placed on the
parameters \cite{PS,DS,PP3}. We are concerned here with a different type
of departure from quantum theory, which is caused by our
insistence on the theory being local.

\section{A local model of collapse}
In the original GRW model, it was proposed that `hits' occurred in a random
fashion, at certain space-time points. The effect of a given hit spread
throughout all space instantaneously. Thus, if we have a single particle
wavefunction $\psi({\bf x},t)$, a hit at the point ${\bf x}_1$, would cause
this to change according to:
\begin{equation}
\psi\to \psi^{\prime}=N \exp \left(-{\beta\over 2}({\bf x}-{\bf x}_1)^2\right)
\psi.
\end{equation}

In order to make this into something that is both local and Lorentz invariant,
we propose instead that a hit at the space-time point ${\rm X_1}\equiv 
({\bf x}_1,t_1)$ only has an effect inside the forward light-cone 
from that point. To ensure Lorentz invariance of 
the hitting function, we must replace the 3-dimensional distance in eq. (1)
by a four-dimensional distance. We cannot use the distance from the 
hitting point to the point on the light-cone since this is identically zero.
Instead, we propose the perpendicular distance from the point on the
light-cone to a four-momentum vector $P_{\mu}$ originating from ${\rm X}$,
where perpendicular is meant in the sense of a Minkowski metric.
With $\psi({\bf x},t)\equiv \psi({\rm X})$, we define this
momentum vector by:
\begin{equation}
{\rm P}_{\mu}^{(1)}=\Re\left({p_{\mu}^{op}
\psi\over \psi}\right)_{{\rm X}={\rm X}_1}.
\end{equation}
If the particle is in an eigenstate of momentum, then this formula will just 
give the four-momentum of the particle. More generally it is the
4-vector form of the particle momentum used in the Bohm hidden-variable model.

\begin{figure}[t]
\begin{center}
\mbox{\psfig{file=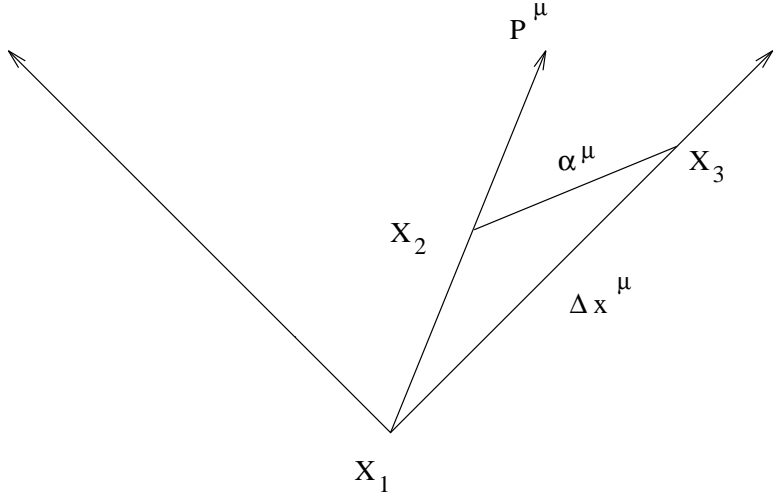,angle=-90}}
\end{center}
\caption{Constructing a Lorentz invariant distance}
\label{Fig:fig1}
\end{figure}
If we denote the vector from the light-cone to ${\rm P}^{\mu}$ by 
$\alpha^{\mu}$ (see Fig. 1) then
the condition that it is perpendicular to ${\rm P}^{\mu}$ is
\begin{equation}
{\rm P}_{\mu}{\rm \alpha}^{\mu}=0.
\end{equation}
The path from ${\rm X}_1$ to ${\rm X}_3$ can be traversed in two ways,
giving another condition
\begin{equation}
k{\rm P}^{\mu}+{\rm \alpha}^{\mu}=\Delta {\rm x}^{\mu},
\end{equation}
for some $k$.
These two equations enable us to find the value of ${\rm \alpha}_{\mu}
{\rm \alpha}^{\mu}$. 
From eq. (2), and using eq. (3), we have 
\begin{equation}
k{\rm P}_{\mu}{\rm P}^{\mu}=
{\rm P}_{\mu}\Delta {\rm x}^{\mu},
\end{equation}
and
\begin{equation} 
{\rm \alpha}_{\mu}{\rm \alpha}^{\mu} =  {\rm \alpha}_{\mu}
\Delta {\rm x}^{\mu}. 
\end{equation}
Also, since $\Delta {\rm x}^{\mu}$ is a null vector, eq. (4) gives
\begin{equation}
k \Delta {\rm x}_{\mu}{\rm P}^{\mu}+ \Delta {\rm x}_{\mu}{\rm \alpha}^{\mu}= 0.
\end{equation}

We can rearrange these three equations to eliminate $k$ and, putting
${\rm P}_{\mu}{\rm P}^{\mu}=m^2c^2$, we have
\begin{equation}
\alpha^{\mu}=\Delta {\rm x}^{\mu}-{1\over m^2c^2}{\rm P}^{\mu}
\left({\rm P}_{\nu}\Delta {\rm x}^{\nu}\right),
\end{equation}
and
\begin{equation}
\alpha_{\mu}\alpha^{\mu}= -{1\over m^2c^2}\left(
{\rm P}_{\mu}\Delta {\rm x}^{\mu}\right)^2.
\end{equation}
This reduces to $\alpha_{\mu}\alpha^{\mu}=-(\Delta {\bf x})^2$ in the
rest frame of the particle, ${\rm P}_{\mu}=(mc,{\bf 0})$.

We therefore postulate that the collapse takes effect along 
the forward light cone from 
${\rm X}_1$, according to
\begin{equation}
\psi_{H_1}({\rm X})=\exp\left(-{\beta\over 2m^2c^2}\left({\rm P}_{\mu}^{(1)}
({\rm X}^{\mu}-{\rm X}_1^{\mu})\right)^2\right)\psi({\rm X}).
\end{equation}
This is our local analogue of eq. (1).

In what follows, we shall simplify the discussion by constraining
the particle to a single spatial dimension (z).
Ideally, we should take the wavefunction to be a solution of the Dirac
equation. However, we wish not to be concerned with any Dirac bispinor,
as the collapse process does not act on the space of spins.
For a free particle, we can instead take the wavefunction to be a solution
of the Klein-Gordon equation. We shall work with a single momentum for which
the initial wavefunction is
\begin{equation}
\psi_0(t,z)=N\exp \left(-iEt+ipz\right),
\end{equation}
where $N$ is some normalization factor.

Given that the forward light-cone is the boundary under consideration, it
is sensible to use light-cone coordinates, $x_{+}=ct+z$, $x_{-}=ct-z$.
The Klein-Gordon equation in this coordinate system reads
\begin{equation}
{\partial^2\over \partial x_{+}\partial x_{-}}\psi
=-{1\over 4}\left({mc\over \hbar}\right)^2\psi.
\end{equation}

Then, if we choose the origin to be at the point of collapse, the
boundary conditions in a general frame of reference are
\begin{eqnarray}
\psi(x_{+},0)&=&N\exp\left(-{i\over 2}(E-p)x_{+}\right)\,
\exp\left(-{\beta\over 8m^2}(E-p)^2 x_{+}^2\right) \\
\psi(0,x_{-})&=&N\exp\left(-{i\over 2}(E+p)x_{-}\right)\,
\exp\left(-{\beta\over 8m^2}(E+p)^2 x_{-}^2\right).
\end{eqnarray}

The solution of the Klein-Gordon equation inside the forward light-cone from
the point of collapse is uniquely defined by these 
boundary conditions.
In order to be able to write this down in a simple form, we shall ignore the 
quantum evolution, i.e. assume ${\hbar\over mc}$ is very small.
For simplicity, we work in the rest frame, in which $p=0$.
Then we can write
\begin{equation}
\psi(x_{+},x_{-})=N\exp\left(-{im\over 2}(x_{+}+x_{-})\right)
w(x_{+},x_{-}).
\end{equation}
Substituting this expression into the Klein-Gordon equation, we have
\begin{equation}
{\partial w\over \partial x_{+}}+{\partial w\over \partial x_{-}}=
-{2i\hbar\over mc}{\partial^2 w\over \partial x_{+}\partial x_{-}},
\end{equation}
where we have included all constants which had been previously set to unity.
The right-hand-side is responsible for the quantum evolution. It can be
treated as a perturbation \cite{CJD}. Here we shall ignore it and just use the
zeroth order solution which is
$w(x_{+},x_{-})=h(x_{+}-x_{-})$. Substituting in the boundary conditions leads
us to a solution:
\begin{equation}
\psi(t,z)=N\exp(-imt) \exp \left(-{\beta\over 2}z^2\right),
\end{equation}
within the forward light-cone of ${\rm X}$.
Outside of this region, the original free-particle solution holds.

If we take the initial wavefunction to be a gaussian with a large spread,
$\psi_0\sim \exp\left(-{z^2\over a^2}\right)$, with $a\gg{\hbar\over mc}$, 
then the momentum states contributing will have
$p\,<{\hbar\over a}\,\ll mc$. We should note that using the collapse radius
for $a$ here gives $p<2\times 10^{-9} mc$, and we are justified in taking this
to have a single momentum component.

To summarise this section, the effect of a single collapse on a single particle
is the same as in the non-relativistic case, except that the effect is only
felt within the forward light-cone of the point of origin of the 
collapse, ${\rm X}_1$.

\section{Single particle affected by two collapses}
A major difference between our local collapse model and that of GRW is
that two independent collapses can occur at space-like separations so that
neither collapse `knows about' the other.
There is no problem with consistency until we arrive at the intersection
of the light-cones arising from the two collapse centers. 
In the region formed by the forward light-cone 
originating from the point of intersection both collapses will be felt,
and we need to define precisely how this happens.

We shall take the two collapses to occur at the points ${\rm X_1}\equiv
(z_1,t_1)\equiv (x_{1+},x_{1-})$ and ${\rm X_2}\equiv(z_2,t_2)
\equiv (x_{2+},x_{2-})$, see Fig. 2.
Before the point of intersection of the two light-cones, the wavefunction
in the regions $w_1$ and $w_2$ will be calculated as in the previous section.
After we reach the intersection point, ${\rm X_3}\equiv(x_{2+},x_{1-})$,
we shall solve the differential equation again with new
boundary conditions along this third light-cone.
\begin{figure}[t]
\begin{center}
\mbox{\psfig{file=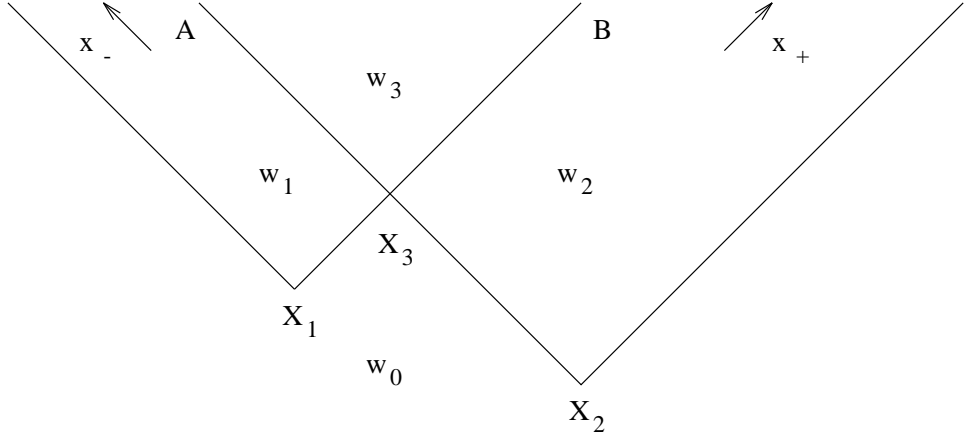,angle=-90}}
\end{center}
\caption{Wavefunction affected by two collapses}
\label{Fig:fig2}
\end{figure}

The boundary conditions are formed, as indeed they were before, by taking
the wavefunction outside of the light-cone, and multiplying by the collapse
factor which arises from the collapse along that particular light-cone.
Thus, for the boundary
condition along ${\rm X_3}A$ we take the wavefunction in the region $w_1$,
which is $\psi_{w_1}(z,t)$, and multiply by the collapse factor arising
from the collapse at ${\rm X}_2$ along ${\rm X}_2 A$. Hence
\begin{equation}
\psi_{{\rm X_1 X_2}}(z,t)=
\exp\left(-{\beta\over 2}\left(\alpha_{{\rm X_2}}\right)^2\right)
\psi_{w_1}(z,t),
\end{equation}
and simliarly along ${\rm X_3}B$
\begin{equation}
\psi_{{\rm X_2 X_1}}(z,t)=
\exp\left(-{\beta\over 2}\left(\alpha_{{\rm X_1}}\right)^2\right)
\psi_{w_2}(z,t),
\end{equation}
where $\alpha_{\rm X_1}$ and $\alpha_{\rm X_2}$ are the perpendicular 
four-distances from the momentum vectors arising from the collapses at
${\rm X_1}$ and ${\rm X_2}$ respectively.

In general, for an arbitrary initial wavefunctions the momentum vectors
arising from each point will be different. Even ignoring the quantum 
evolution, this could lead to solutions of the
differential equation which are quite complicated. For simplicity,
we shall deal
with the case when the momenta arising from each collapse center are equal.
In this situation, we can again work in the frame where ${\bf p}=0$.
This means that the boundary conditions are (having extracted the plane-wave
term and normalization as before):
\begin{equation}
w(x_{2+},x_{-}) =  \exp\left(-{\beta\over 2}(z-z_1)^2\right)
\exp\left(-{\beta\over 8}(x_{-}-x_{2-})^2\right),
\end{equation}
along ${\rm X}_3 A$, and
\begin{equation}
w(x_{+},x_{1-})= \exp\left(-{\beta\over 2}(z-z_2)^2\right)
\exp\left(-{\beta\over 8}(x_{+}-x_{1+})^2\right),
\end{equation}
along ${\rm X}_3 B$.

We are only interested in the zeroth order solution to the 
differential equation, eq. (16), so of course we have
$w(x_{+},x_{-})= h(x_{+}-x_{-})$, as before.
Substituting the boundary conditions, we find that
\begin{equation}
h(x_{+}-x_{1-})=\exp\left(-{\beta\over 8}\left[(x_{+}-x_{1-}-x_{2+}+x_{2-})^2
+(x_{+}-x_{1+})^2\right]\right),
\end{equation}
which may be rewritten as
\begin{equation}
h(\lambda)=\exp\left(-{\beta\over 8}\left[(\lambda-x_{2+}+x_{2-})^2
+(\lambda-x_{1+}+x_{1-})^2\right]\right),
\end{equation}
leading to a wavefunction
\begin{equation}
\psi(t,z)=N(0,m)\exp\left(-i{mc^2\over \hbar}t\right)
\exp\left(-{\beta\over 2}(z-z_1)^2\right)
\exp\left(-{\beta\over 2}(z-z_2)^2\right).
\end{equation}

Here we find that, in this case, the two collapses are equivalent to a single
collapse at the point ${1\over 2}({\rm X_1}+{\rm X_2})$, 
but with twice the `strength'. This is 
certainly what would be expected in the non-relativistic limit if we were
to have two collapses, although this solution only holds in the forward
light-cone of the intersection point ${\rm X_3}$. It should be noted that
in the non-relativistic situation, the wavefunction at the point of the second
collapse would have already been reduced by the first, so the probability of
the second collapse occurring would be very small.

We now briefly consider the question of the order of the two collapses.
Take the situation shown in fig. 3, where the collapse at ${\rm X_1}$
happens later than the collapse at ${\rm X_2}$, in the frame in which $p=0$.
The relative size of the two peaks depends on the distances from
${\rm X_1}$ to ${\rm A_2}$ and ${\rm X_2}$ to ${\rm A_1}$. As can be seen
from the diagram, these distances are of course equal, so the exponentials
by which we multiply the two wavefunctions will be equal, at the two peaks, 
and thus as the wavefunction is not time-dependent, the two peaks will have the
same size. The order of the collapses is immaterial when we have a single 
momentum component.

\begin{figure}[t]
\begin{center}
\mbox{\psfig{file=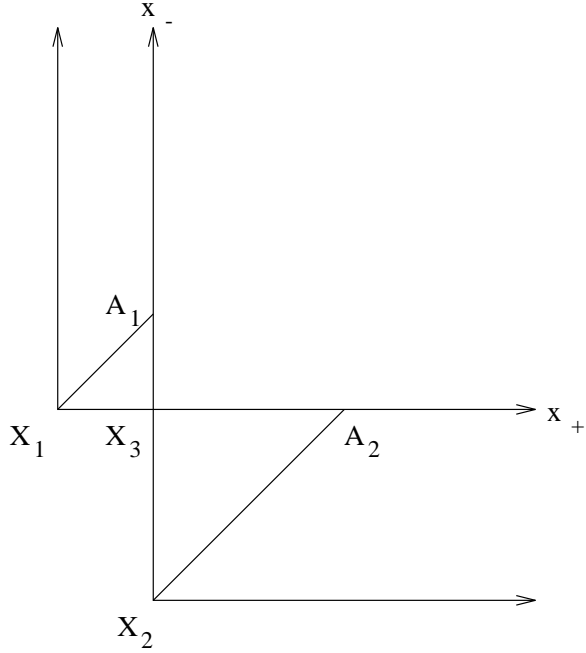,angle=-90}}
\end{center}
\caption{The effect of the relative times of the collapses}
\label{Fig:fig3}
\end{figure}

\subsection{Superposition of Two Wavepackets}

We now want to examine a typical measurement situation, where the
initial wavefunction is a sum of two well-separated peaks, e.g.
\begin{equation}
\psi_0(z)=N\left[\exp\left(-\alpha(z-z_1)^2\right)
+\exp\left(-\alpha(z-z_2)^2\right)\right],
\end{equation}
with $|z_1-z_2|\gg {1\over \sqrt \alpha}$. So any collapse 
which occurs will, with probability essentially one,
be centered around one of the two peaks of the wavefunction.
An important property of the GRW-type models which we retain 
is that the probability of
a collapse occurring at a point ${\bf x}$ is proportional to 
$|\psi({\bf x})|^2$; 
with a single collapse, the wavefunction will be reduced to a single peak
in a time $t\approx {|z_1-z_2|\over c}$.

However, as before, the relativistic model allows the possibility of there
being two collapse events, one centered on each peak, providing that each 
collapse event is outside of the forward light-cone of the other.
On a constant time slice, 
there may persist peaks in each region, but we are predominately concerned with
the shape of the wavefunction at later times, i.e. after the
intersection of the two forward light-cones.

We assume that the momentum vectors defined at the two collapse points
are (to a sufficiently good approximation) the same, so that we can 
again work in the frame for which ${\bf p}=0$. Then, with
the same approximations as before, the final state wavefunction will
just be the initial state wavefunction multiplied by the collapse functions 
arising from the spatially separated collapses. This final state
wavefunction can be written as
\begin{equation}
\psi_f=N\left[\exp\left(-(\alpha+\beta)(z-z^{\prime}_1)^2\right)
+\exp\left(-(\alpha+\beta)(z-z^{\prime}_2)^2\right)\right],
\end{equation}
and we again have the two peaks, only now their centers have been shifted,
according to
\begin{eqnarray}
z^{\prime}_1&=& z_1+{{1\over 2}\beta\over \alpha+\beta}(z_2-z_1) \\
z^{\prime}_2&=& z_2+{{1\over 2}\beta\over \alpha+\beta}(z_1-z_2).
\end{eqnarray}

Obviously if the peaks were very sharp in the initial wavefunction,
then the shift will 
be quite small. However, it is certainly possible that the shift will be 
sufficient for the new peak to lie well into the tail of the initial peak, 
where it would have been extremely unlikely that the particle could be found.
This will be the case if
$\exp\left(-{\alpha\beta^2\over 4(\alpha+\beta)^2}(z_2-z_1)^2\right)\ll 1$.
For a pair of sharp initial peaks, and $\alpha\gg \beta$, this reads
$|z_2-z_1|\gg{2\sqrt \alpha\over \beta}$.

In general we might expect the localization of the two peaks to be less
than, but of the order of, the GRW collapse size, i.e. $\beta< \alpha$ but
of the same order. This means that the peaks are shifted by something 
around ${1\over 4}$ of their separation.

\section{The Born Probability Rule}

In orthodox quantum theory, the probability that a measurement outcome will
correspond to one of 2 peaks is proportional to the square integral of the
weight of each peak.
The same result holds in GRW, where it is a consequence of the probability 
rule for a hit occurring at a particular point \cite{GRW}.
Here we shall again guarantee this result,
{\it for a single collapse}, by postulating that the 
probability of this collapse occurring at one of the peaks is 
proportional to the integral of the square of the wavefunction over the 
peak. Thus with an initial state (in the rest frame):
\begin{equation}
\psi_0(z)=N\left[a \exp\left(-\alpha(z-z_1)^2\right)
+b \exp\left(-\alpha(z-z_2)^2\right)\right],
\end{equation}
with $|a|^2+|b|^2=1$, the collapse will occur near $z_1$ or $z_2$ in the 
ratio of $|a|^2$ to $|b|^2$.

However, we now have to consider carefully the possibility of more than one
collapse occurring. This means of course that both peaks can change
their magnitudes. We take account of this by allowing $a$ and $b$ in eq. (30)
to be functions of time. Consistent with the requirement of a {\it local}
model we postulate that the probability of a collapse at $z_1$ at time
$t_1$ is proportional to ${|a(t)|^2\over |a(t)|^2+|b(t_R)|^2}$, where 
$t_R=t_1-{|z_1-z_2|\over c}$, the retarded time.

The probability of a particular peak persisting depends of course on the
number of collapses that occur, and the time taken
for the signal of a collapse to reach the other peak($T$).
Here we shall evaluate the probability of
peak 1 dominating.
There will be contributions to this 
probability from all possible numbers of collapses. We shall assume that 
$\lambda T\ll 1$, and so make an expansion in this parameter. We shall 
calculate the first three terms in this series, 
i.e. work to order $(\lambda T)^2$.

\begin{figure}[pt]
\begin{center}
\mbox{\psfig{file=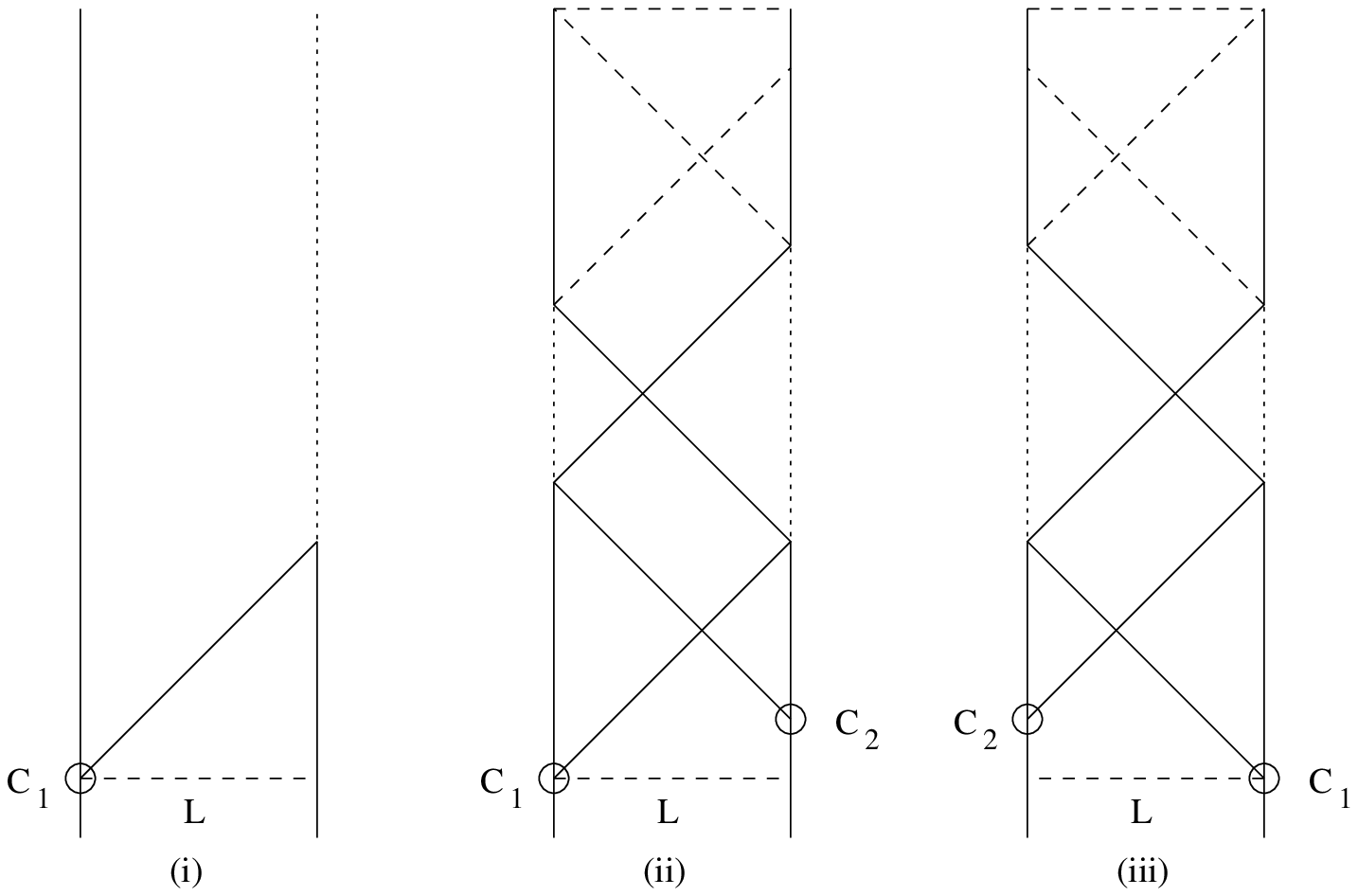,angle=-90}}
\vspace{0.5cm}

\mbox{\psfig{file=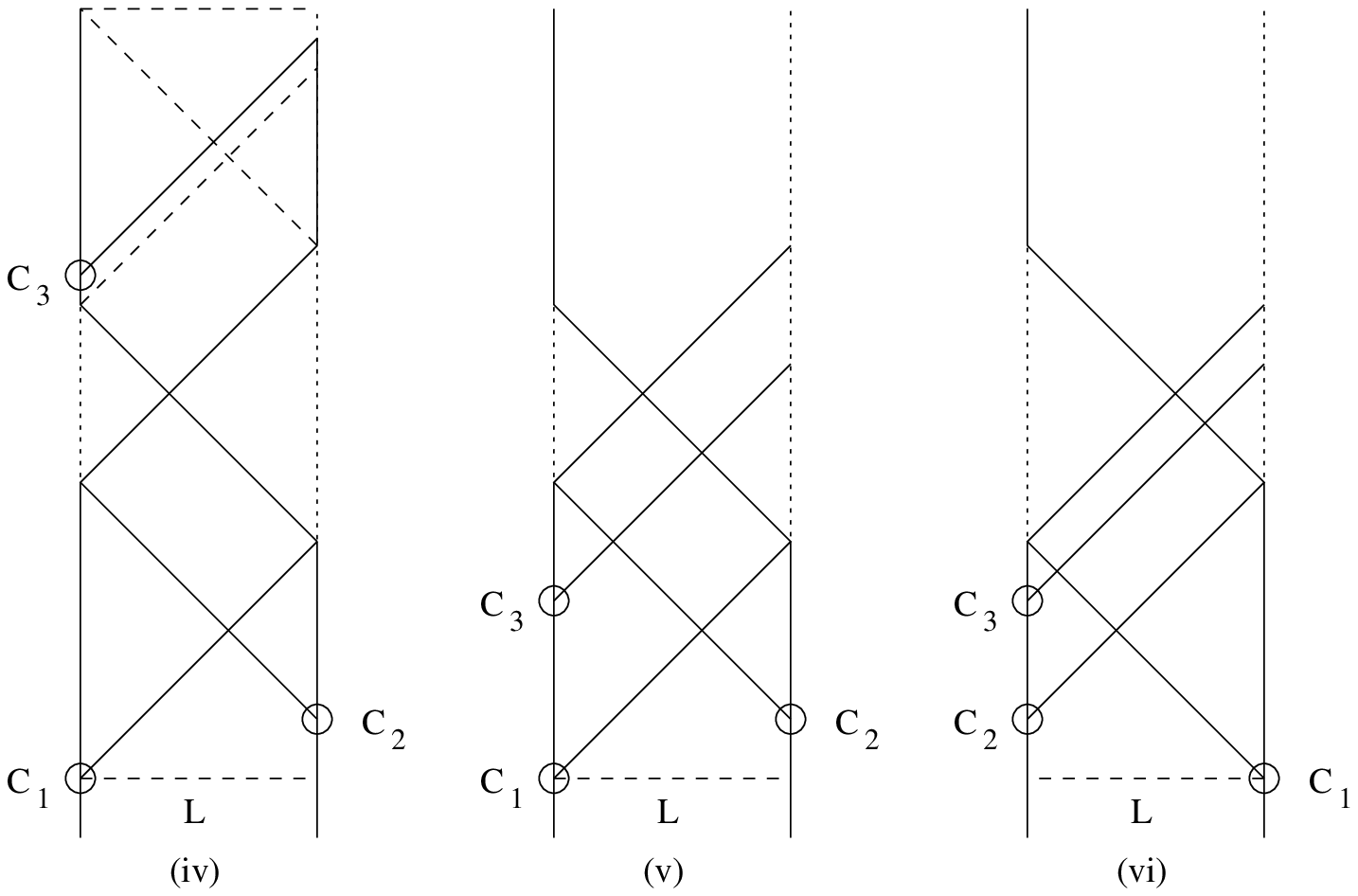,angle=-90}}
\end{center}
\end{figure}

\begin{figure}[t]
\begin{center}
\mbox{\psfig{file=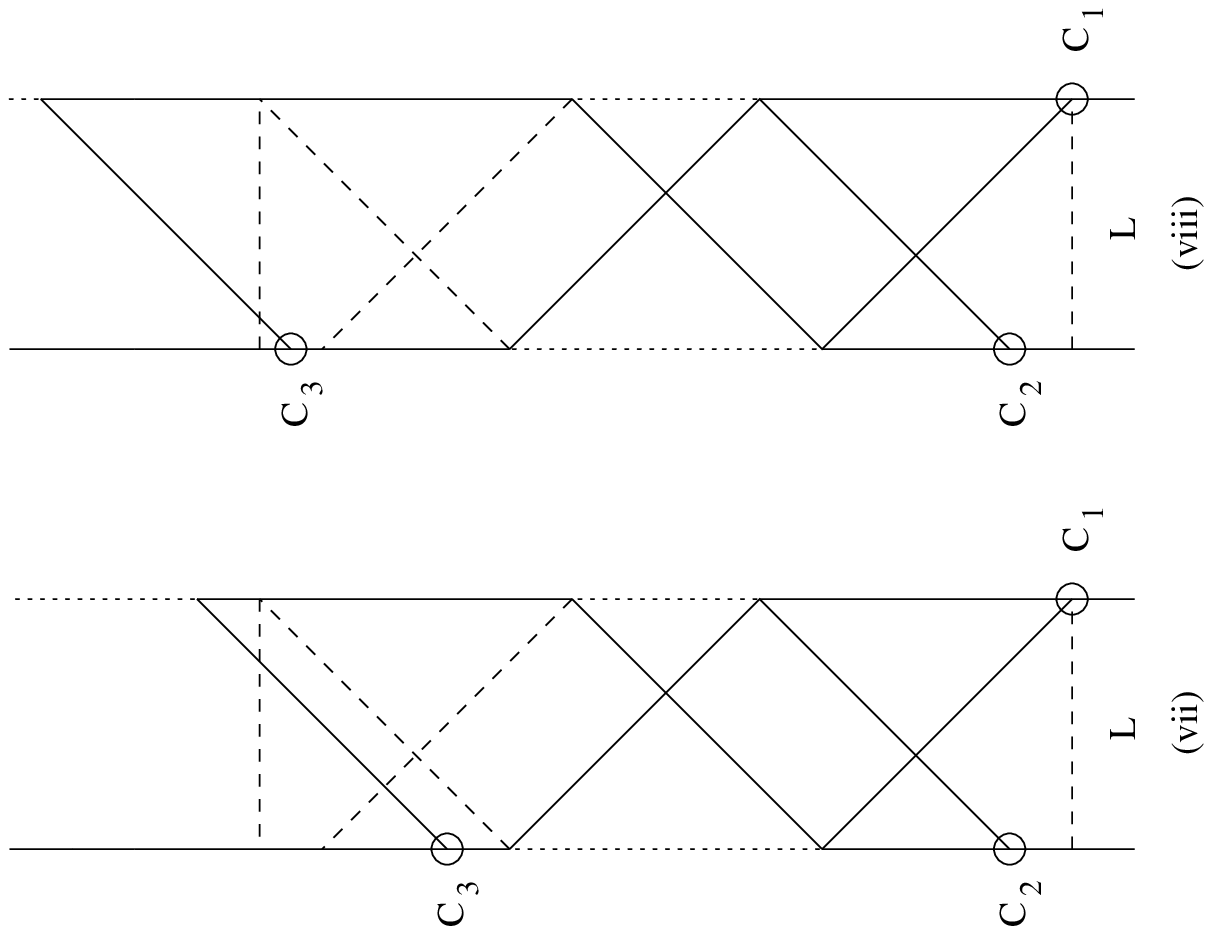,angle=-90}}
\end{center}
\caption{Collapse processes which contribute to second order in $\lambda T$.}
\label{Fig:fig4}
\end{figure}

The collapse processes which contribute to this order are illustrated in 
Fig. 4.
The separation of the two peaks is $z_1-z_2\equiv L\equiv Tc$. In order to
assess the probability of a collapse occuring on a particular peak, we
look at the relative sizes of the peaks along the backward light-cone.
In the figure, the solid vertical lines indicate where a collapse on the peak
is possible, whereas the dotted lines indicate that a collapse is not possible.
When a collapse is deemed possible, the probabilities for each side will
be $|a|^2$ and $|b|^2$ respectively, if both can occur, or 1 and 0 if only
one of these is possible.
We need to calculate the probability contributions from each diagram
separately.
\vspace{.5cm}

\noindent{\bf Diagram (i)}
In this case a single collapse is successful. There are no collapses on the
other peak before it has received the signal from the first collapse.
The probability of this occurring is
\begin{equation}
P_i=|a|^2\exp\left(-\lambda T |b|^2\right)=|a|^2\left(1-\lambda T|b|^2
+{1\over 2}(\lambda T)^2|b|^4+O([\lambda T]^3)\right).
\end{equation}

\noindent{\bf Diagram (ii)}
Here we have two specified collapses, $C_1$ and $C_2$,
one centered on each peak, with that on
the peak number 1 occurring first, and with no further collapses 
before the time indicated
by the dashed line.
The probability can be written as
\begin{eqnarray}
P_{ii}&=&|a|^2\int^T_0 \exp\left(-\lambda T |b|^2\right)\lambda dt |b|^2
\exp\left( -\lambda (T+t)|a|^2\right)\exp\left(-2\lambda T\right) 
\exp\left(-\lambda t|b|^2\right) P
\nonumber \\ & = &
|a|^2|b|^2(\lambda T)[1-{7\over 2}(\lambda T)]\, P +O([\lambda T]^3),
\end{eqnarray}
where $P$ is the overall probability of peak 1 dominating.

\noindent{\bf Diagram (iii)}
This diagram is a mirror image of diagram (ii), and the probability 
contribution is in fact the same. 
\begin{equation}
P_{iii}=P_{ii}.
\end{equation}

\noindent{\bf Diagram (iv)}
The initial collapses are the same as in diagram (ii), only we have a
further collapse occurring before the dashed line. (n.b. if there are no 
further collapses before this time, then we essentially have the situation 
with which we started.)
This added collapse gives rise to another integral in the calculation
of the probability
\begin{eqnarray}
P_{iv}&=&\begin{array}[t]{lr}
\multicolumn{2}{l}{\displaystyle{
|a|^2\int^T_0 \exp\left(-\lambda T|b|^2\right)\lambda dt|b|^2
\exp\left(-\lambda (T+t)|a|^2\right)}} \\
\multicolumn{2}{r}{\displaystyle{\hspace{2cm}\times
\int^{T+t}_0 \exp\left(-2\lambda T\right)
\lambda dt^{\prime}\exp\left(-\lambda t^{\prime}|b|^2\right)}}
\end{array}\nonumber \\
&=& {3\over 2}(\lambda T)^2|a|^2|b|^2 +O([\lambda T]^3).
\end{eqnarray}

\noindent{\bf Diagram (v)}
This time we have two collapses centered on peak 1 before the collapse on
the second peak has taken full effect, with the first collapse centered on 
peak 1.
\begin{eqnarray}
P_v&=&|a|^2\int^T_0 \exp\left(-\lambda T |b|^2\right) \lambda dt |b|^2
\lambda |a|^2 (T+t) \exp \left(-\lambda (T+t)|a|^2\right) \nonumber \\
& = & {3\over 2}(\lambda T)^2|a|^4|b|^2 +O([\lambda T]^3).
\end{eqnarray}

\noindent{\bf Diagram (vi)}
This is similar to the previous diagram, with the solitary collapse on
peak number 2 occurring first.
\begin{eqnarray}
P_{vi}&=&|b|^2\int^T_0\exp\left(-\lambda T|a|^2\right)\lambda dt |a|^2 
\exp\left(-\lambda (T+t)|b|^2\right)\int^T_t \lambda dt^{\prime}|a|^2
\nonumber \\ & = &
{1\over 2}(\lambda T)^2|a|^4|b|^2 +O([\lambda T]^3).
\end{eqnarray}

\noindent{\bf Diagram (vii)}
This diagram is similar to diagram (iii) in the same way as diagram (iv)
is related to diagram (ii).
\begin{eqnarray}
P_{vii}&=&\begin{array}[t]{lr}\multicolumn{2}{l}{\displaystyle{
|b|^2\int^T_0 \exp\left(-\lambda T|a|^2\right)\lambda dt |a|^2
\exp\left(-\lambda (T+t)|b|^2\right)}}\\ \multicolumn{2}{r}{\displaystyle{
\int^{T-t}_0 \lambda dt^{\prime}
\exp\left(-2\lambda T\right)\exp \left(-\lambda t^{\prime}|b|^2\right)}}
\end{array}\nonumber \\ & = &
{1\over 2}|a|^2|b|^4 (\lambda T)^2 +O([\lambda T]^3).
\end{eqnarray}

\noindent{\bf Diagram (viii)}
This is almost the same as the last diagram, except that the time of the last
collapse gives it a different probability of occurring.
\begin{eqnarray}
P_{viii}&=&\begin{array}[t]{lr}\multicolumn{2}{l}{\displaystyle{
|b|^2\int^T_0 \exp\left(-\lambda T|a|^2\right)\lambda dt |a|^2
\exp\left(-\lambda (T+t)|b|^2\right)}}\\ \multicolumn{2}{r}{\displaystyle{
\hspace{1cm}\int^t_0 \lambda dt^{\prime} |a|^2
\exp\left(-2\lambda T\right)\exp\left(-\lambda t|a|^2\right)
\exp\left(-\lambda(T-t+t^{\prime})|b|^2\right)}}
\end{array} \nonumber \\ & = &
{1\over 2}|a|^4|b|^2(\lambda T)^2 +O([\lambda T]^3).
\end{eqnarray}

Adding all these calculated probabilities together and rearranging,
leads to
\begin{equation}
P=|a|^2+\lambda T|a|^2|b|^2(|a|^2-|b|^2)-{1\over 2}(\lambda T)^2
|a|^2|b|^2(|a|^2-|b|^2)(5-4|a|^2|b|^2).
\end{equation}

If the initial superposition is equally weighted, then as expected the
probabilities for each peak to dominate are equal.
However, if we start with an unequal superposition of the two gaussian peaks,
then in this model the probability of obtaining the 
initially higher peak increases with the peak separation.

\section{Two-particle correlated wavefunction}

We will now deal with the case where we have two particles with a correlated
wavefunction, for instance an EPR-type situation corresponding to the
measurement of the spins of two fermions in a correlated state
\footnote{It should be noted that there is now an ambiguity in the
definition of ${\rm P}_{\mu}$ (see eq. (2)) since the right-hand-side
depends upon the value of the position variable for the other particle. The
simplest procedure would be to say that the collapse selects a random value 
for this, i.e. that it is actually associated with a point in 
configuration space. This issue does not concern us here since we are 
restricting our attention to situations where all `3-momenta' are, to
a sufficiently good approximation, zero in some reference frame.}.
The initial wavefunction can be written in the form:
\begin{eqnarray}
\psi(z_1,z_2)&=&\displaystyle{
N\left[a\phi_1(z_1)\phi_2(z_2)+b\chi_1(z_1)\chi_2(z_2)\right]}
\nonumber \\ &=&
\begin{array}[t]{lr}
\multicolumn{2}{l}{\displaystyle{N\left[
a\exp\left(-\alpha(z_1-z_{11})^2\right)
\exp\left(-\alpha(z_2-z_{21})^2\right)\right.}}\\
\multicolumn{2}{r}{\displaystyle{\left.\hspace{2cm}
+b\exp\left(-\alpha(z_1-z_{12})^2\right)
\exp\left(-\alpha(z_2-z_{22})^2\right)\right],}}
\end{array}
\end{eqnarray}
where $z_{11},z_{12}$ refer to the center of the peaks corresponding to 
particle 1, and similarly for particle 2.
We assume that the two peaks for each particle do not overlap significantly, so
that a collapse centered on one will kill the other peak, i.e.  
$\alpha(z_{11}-z_{12})^2\gg 1$ and $\beta(z_{11}-z_{12})^2\gg 1$. Also, for
simplicity, we will consider only the case when this peak separation itself is
negligible compared with the separation of the two particles, for instance,
$|z_{11}-z_{21}|\gg |z_{11}-z_{12}|$. This of course corresponds to the 
actual experimental situations in tests of the Bell inequality.

Collapse processes centered on each particle will be taken to be independent,
and they will have the same effect on the wavefunction as before. However,
in the case where we have two `incompatible' collapses, the situation will have
changed in that as each collapse acts on a different part of the wavefunction,
they will not `interfere' at any point in space. At a time
after signals from both collapses have reached the other, the wavefunction in
the intermediate region will just be multiplied by the two independent
collapse factors irrespective of the momentum states from which they
were constructed, i.e. for collapses centered at $z_{11}$ and $z_{22}$,
\begin{equation}
\psi^{\prime}=\psi\exp\left(-{\beta\over 2}(z_1-z_{11})^2\right)
\exp\left(-{\beta\over 2}(z_2-z_{22})^2\right).
\end{equation}

As before, we should again ask which part of the wavefunction will dominate.
We can do a similar calculation to before, with the probability of a collapse
occurring on a particular peak being either $|a|^2,|b|^2$, 1 or zero.
However, as the peak separation of either particle is 
considered to be negligible compared to the separation between the
two particles, we shall take the signal of a collapse to travel 
instantaneously to the other peak connected to that particle. For a single 
particle, the probability of another collapse in the actual time taken is very 
small. 

It turns out that the probability for a particular peak to dominate
is actually the same
as when we only have one particle. 

\begin{figure}[pt]
\begin{center}
\mbox{\psfig{file=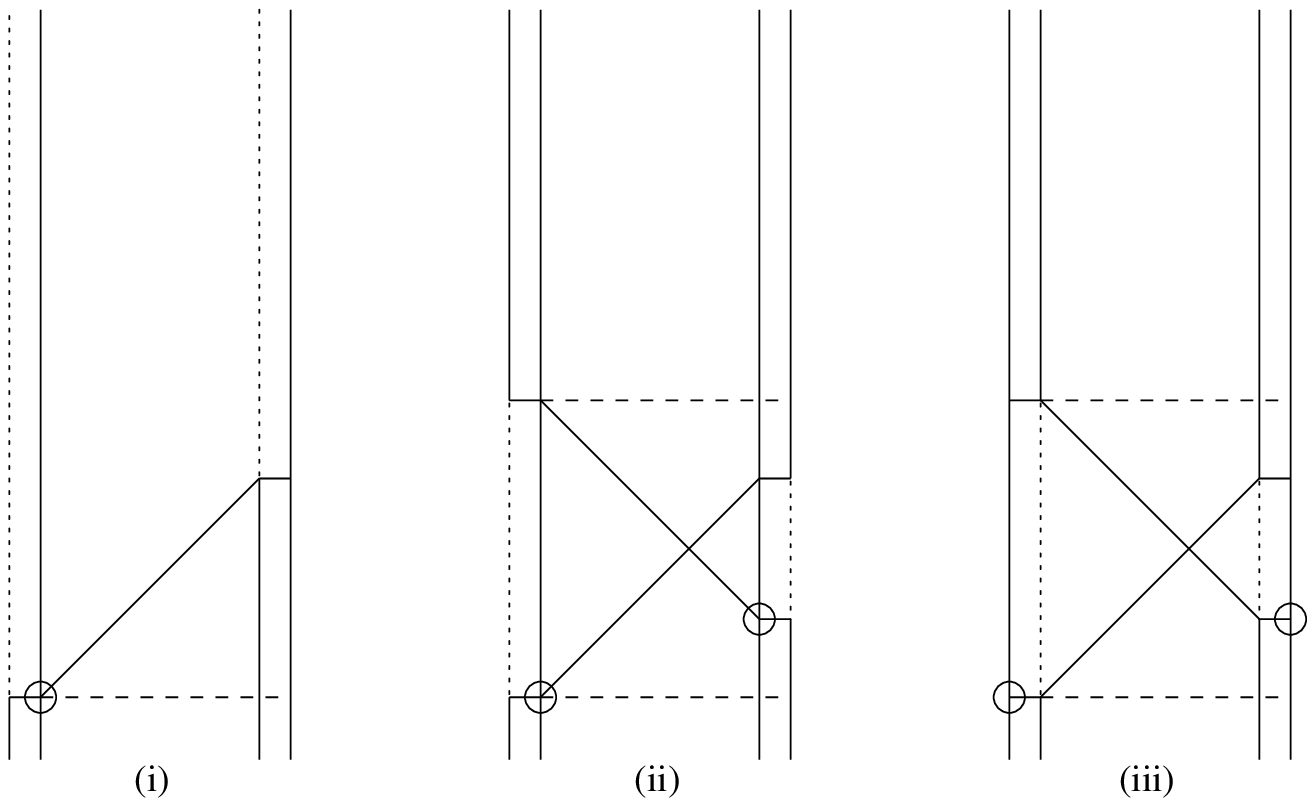,angle=-90}}
\vspace{.5cm}

\mbox{\psfig{file=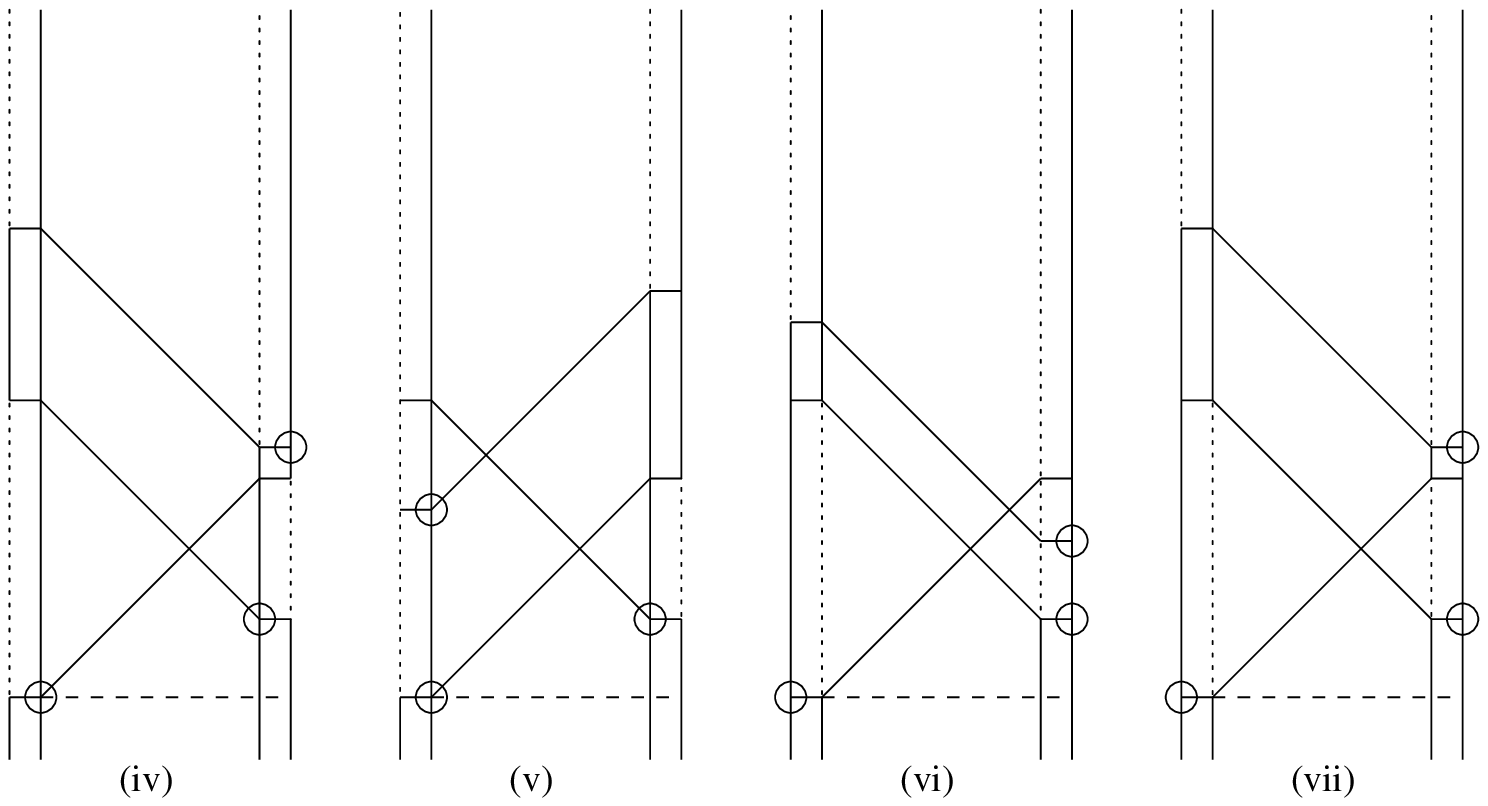,angle=-90}}
\end{center}
\caption{The collapse processes which contribute 
to second order in $\lambda T$ for two correlated particles.}
\label{Fig:fig5}
\end{figure}

Fig. 5 illustrates the diagrams which 
contribute to second order in $\lambda T$,
ignoring the separation between the two peaks on one side. As before the solid
lines indicate that a collapse is deemed possible, whereas a collapse cannot 
occur where the line is dotted. The points where a collapse occurs are 
circled. The peaks for particle 1 are those to the left, with the $\psi$ peak
the right one of these and the $\chi$ peak on the left. We calculate the
probability of the $\psi$ peaks dominating.
\vspace{.5cm}

\noindent{\bf Diagram (i)}
The simplest case where we have no incompatible collapses.
\begin{eqnarray}
P_{2i}&=&|a|^2\left[\exp\left(-\lambda T\right)+
\int^T_0 \lambda |a|^2 dt \exp\left(-\lambda t\right) \right] \nonumber \\ &=&
|a|^2\left(1-\lambda T|b|^2+(\lambda T)^2|b|^2\right)+ O([\lambda T]^3).
\end{eqnarray}

\noindent{\bf Diagram (ii)}
As in the one-particle case, we have two specified collapses, one affecting
each particle, but incompatible.
\begin{eqnarray}
P_{2ii}&=&|a|^2\int^T_0 \lambda |b|^2 dt \exp\left(-2\lambda(T+t)\right)P_2
\nonumber \\ &=& |a|^2|b|^2(\lambda T)\left(1-3(\lambda T)\right)P_2
+O([\lambda T]^3),
\end{eqnarray}
where $P_2$ is the overall probability of the $\psi$ peaks
dominating for this two-particle
wavefunction.
 
\noindent{\bf Diagram (iii)}
This is just the mirror image of the previous diagram, with identical
contribution to the probability.
\begin{equation}
P_{2iii}=P_{2ii}.
\end{equation}

\noindent{\bf Diagram (iv)}
As in diagram (ii) but with an additional collapse prior to both particles
having knowledge of both previous collapses.
\begin{eqnarray}
P_{2iv}&=&|a|^2\int^T_0 \lambda |b|^2 dt \int^t_0 \lambda |a|^2 dt^{\prime}
\exp\left(-\lambda(T+t^{\prime})\right)\exp\left(-\lambda (T-t+t^{\prime})
\right) \nonumber \\
&=& {1\over 2}|a|^4|b|^2(\lambda T)^2.
\end{eqnarray}

\noindent{\bf Diagram (v)}
One particle has two compatible collapses dominating the collapse on the other
particle.
\begin{eqnarray}
P_{2v}&=&|a|^2\int^T_0 \lambda |b|^2 dt \exp\left(-\lambda T\right)
\exp\left(-\lambda (T+t)\right)\int^{T+t}_0\lambda dt^{\prime}
\exp\left(-\lambda t^{\prime}\right) \nonumber \\ &=&
{3\over 2}|a|^2|b|^2(\lambda T)^2.
\end{eqnarray}

\noindent{\bf Diagram (vi)}
As in the last diagram, but the order of the first two collapses is reversed.
\begin{eqnarray}
P_{2vi}&=&|b|^2\int^T_0\lambda |a|^2 dt \exp\left(-\lambda T\right)
\exp\left(-\lambda(T+t)\right)
\int^{T-t}_0\lambda dt^{\prime}\exp\left(-\lambda t^{\prime}\right)
\nonumber \\ &=& {1\over 2}|a|^2|b|^2(\lambda T)^2.
\end{eqnarray}

\noindent{\bf Diagram (vii)}
Similar to diagram (iii) with the same difference as between diagrams
(ii) and (iv).
\begin{eqnarray}
P_{2vii}&=&|b|^2\int^T_0\lambda |a|^2 dt \exp\left(-\lambda T\right)
\exp\left(-\lambda (T+t)\right)\int^t_0\lambda |a|^2 dt^{\prime}
\exp\left(-\lambda (T-t+t^{\prime})\right) \nonumber \\ &=&
{1\over 2}|a|^4|b|^2 (\lambda T)^2.
\end{eqnarray}

Adding these probabilities together, gives
\begin{equation}
P_2=|a|^2+\lambda T|a|^2|b|^2(|a|^2-|b|^2)-{1\over 2}(\lambda T)^2
|a|^2|b|^2(|a|^2-|b|^2)(5-4|a|^2|b|^2),
\end{equation}
which is the same probability as that obtained for the one-particle,
two-peak wavefunction.

\section{Magnitudes}

The crucial parameter in the above discussion is
\begin{equation}
\lambda T={L\over c}{N\over \tau_{col}},
\end{equation}
where $\tau_{col}$ is the collapse time for a single particle and $N$ is
the number of particles involved in the measurement apparatus. If we take
$L=10 \,{\rm m}$ corresponding to roughly the largest separation in the Aspect
experiments \cite{ASP}, and use the GRW value ($10^{16} \,{\rm s}$) for
$\tau_{col}$ this becomes
\begin{equation}
\lambda T={N\over 3}\times 10^{-23}.
\end{equation}
It is clear that with a macroscopic apparatus this number could well be of
the order of unity or larger. Hence, the possibility of detecting
violations of the quantum probability rule certainly exist. On the other hand,
the uncertainty in the value of the parameter $\tau_{col}$, and the possibility
of variation in the precise predictions of particular versions of the collapse
models, e.g. as discussed in \cite{PS,DS,PP3} etc., mean that it is not 
possible to rule out our local form of collapse models.
In order to do this, or to see the new effects they predict, it would
be necessary to do experiments in which $L$ and $N$ are as large as
possible, and in which the ratio of $|a|$ to $|b|$ in the measured state
lies in the middle of the range [0,1]. It would of course also be
necessary to do a careful analysis of the actual measuring apparatus, rather
than just modelling it by a `pointer' as in \S 5 above.

It should be noted that the only necessary constraint on the collapse time of 
the apparatus (${\tau_{col}\over N}$) is that it is less than the time of
perception ($\tau_{per}$), which is certainly not less 
than $10^{-4}\, {\rm s}$. Hence we know $\lambda T>
{T\over \tau_{per}}\sim
10^{-3}$. There is ample space here for values of $\lambda T$ 
considerably less than unity, for which deviations from the
quantum probability law would not have been seen.
On the other hand, by using larger apparatus and/or larger values of $L$, the
effects should certainly become observable.

\section{Summary}

An important contribution of the original GRW model was that it showed 
the {\it possibility} of defining a precise model in which collapse happens
as a physical process in such a way that the tested predictions of
quantum theory still held (This is independent of the issue of whether 
nature actually {\it chooses} this particular solution of the 
measurement problem).

In the same spirit we have shown here the possibility of defining a precise
collapse model which is {\it local} and {\it Lorentz-invariant}. The issue
of whether it is still consistent with all experiments is somewhat less
clear, but as stated above the freedom in the choice of
parameters almost certainly means that it can be made consistent. To this
extent we have shown that the widespread belief that Bell's theorem combined
with the results of the Aspect et al. experiments \cite{ASP} mean that
any realistic model of quantum theory (apart from `many-worlds' versions)
must be explicitly non-local, is false.

\end{document}